\documentclass{pasa}%

\usepackage{graphicx}
\usepackage{amsmath}
\usepackage[utf8]{inputenc}
\usepackage[font=scriptsize]{caption}

\title{The effect of quantum fluctuations in compact star observables}
\author[P. P\'osfay et al.]{P. P\'osfay$^{1,2}$, G. G. Barnaf\"oldi$^1$ and A. Jakov\'ac$^2$
\affil{$^1$Wigner Research Centre for Physics of the H.A.S., P.O. Box, H-1525 Budapest, Hungary}%
\affil{$^2$Institute of Physics, E\"otv\"os University, 1/A P\'azm\'any P. S\'et\'any, H-1117 Budapest, Hungary}
}%

\jid{PASA}
\doi{10.1017/pas.\the\year}
\jyear{\the\year}

\usepackage{aas_macros}
\usepackage{hyperref} 
\hypersetup{colorlinks,citecolor=blue,linkcolor=blue,urlcolor=blue}

\newcommand{\dd}{ {\mathrm d} } 

\newcommand{\ph}{\varphi}

\newlength{\szovszel}
\newlength{\slashszel} 
\newcommand*{\sls}[1]{\mbox{%
    \settowidth{\szovszel}{\ensuremath{#1}}%
    \settowidth{\slashszel}{\ensuremath{\slash}}%
    \hspace*{0.5\szovszel}%
    \hspace*{-0.5\slashszel}%
    \slash%
    \hspace*{-0.5\szovszel}%
    \hspace*{-0.5\slashszel}%
    \ensuremath{#1}%
  }}

\renewcommand{\d}{\partial}

\usepackage{xcolor}

\begin{document}

\begin{frontmatter}
\maketitle

\begin{abstract}
Astrophysical measurements regarding compact stars are just ahead of a big evolution jump, since the NICER experiment deployed on ISS on 14\textsuperscript{th} June 2017. This will provide soon data that would enable the determination of compact star radius with less than $10\,\%$ error. This can be further constrained by the new observation of gravitational waves originated from merging neutron stars, GW170817. This poses new challenges to nuclear models aiming to explain the structure of super dense nuclear matter found in neutron stars.

A detailed studies of the QCD phase diagram shows the importance of bosonic quantum fluctuations in the cold dense matter equation of state. 
Here we used a demonstrative model with  one bosonic and one fermionic degree of freedom coupled by Yukawa coupling we show the effect of bosonic quantum fluctuations on compact star observables such as mass, radius and compactness. 

We have also calculated the difference in the value of compressibility which is caused by quantum fluctuations. The above mentioned quantities are  calculated in mean field, one-loop and in high order many loop approximation. The results show that the magnitude of these effects is in the range of $4\text{-}5 \, \%$, which place it into the region where modern measurements may detect it.  This forms a base for further investigations that how these results carry over to more complicated models. 
\end{abstract}


\begin{keywords}
Dense matter -- Stars: neutron -- Equation of state -- Astroparticle physics -- Gravitational waves
\end{keywords}
\end{frontmatter}

\section{INTRODUCTION }
\label{sec:intro}

Studying the interior of compact stars is challenged by the lack of direct observations.
However, astrophysical measurements regarding these celestial objects are just ahead of a big evolution jump, since the NICER experiment was deployed on ISS on 14\textsuperscript{th} June 2017~\cite{nasanicer}. The improved X-ray data analysis methods by~\cite{Ozel:2016} provide more and more precise data, which enable us to determinate the radii of compact stars with less than $10\,\%$ error~\cite{Ozel:2015ykl}. Moreover, a further window to the sky, the discovery of gravitational waves~\cite{gw:1,gw:2} introduced a novel method to investigate the inner structures of neutron stars in a completely new manner as presented in~\cite{Rezzolla:2016nxn}. The combined measurements of gravitational and electromagnetic observations, GW170817 by~\cite{TheLIGOScientific:2017qsa} highlights the era of multi messenger astronomy which can provide higher accuracy compact star parameters. 

Despite the ever increasing accuracy of astrophysical measurements, the modeling of the nuclear matter using neutron star data is still plagued with the \textit{masquarade} problem: different equation of states (EoS) yield similar values for observables of compact astrophysical objects~\cite{Alford:2004}. This means that any method which can support us to select models considering compact star observables, can resolve the masquerade problem and provide constraints on the models of the superdense nuclear matter.

The high-energy accelerator experiments and the lattice QCD calculations are proved to be a valuable tool for studying the properties of nuclear matter in states characterized by high temperature ($T\approx \,150\, \text{MeV}$) and low densities. Still, since the matter of compact stars lie at the high-density and cold ($T < \,10\, \text{MeV}$) region of the phase diagram of the strongly interacting matter, it  is still unreachable by these techniques above. Here, the phenomenological, effective, and non-perturbative field theory models are usually applied and the state-of-the-art equation of state calculations must contain all the symmetries of the field theory including the relativistic and quantum effects (fluctuations) as well. These theoretical approaches are the only tool so far to connect the astrophysical observations with the nuclear theory description of the neutron star matter. One of the aim of the recent theoretical developments is to provide a mathematical method to handle approximations, while keeping the physically relevant information (e.g. phase transition, degrees of freedom, bound states, {\sl etc}). 

Detailed studies of the QCD phase diagram shows the importance of correct treatment of bosonic fluctuations~\cite{Zsolt:2007}. Considering quantum fluctuations in effective theories of nuclear matter may help us differentiate between models, since the effect of quantum fluctuations may vary model by model. The first step in this investigation is to show that taking into account quantum fluctuations change the physical predictions of the model. In this case the observable quantities of compact stars are expected to change too. 

In this work we use the results of Refs. \cite{Posfay:2015} and~\cite{harmonic:2017,sqmproc:2017} and apply a simple model including a Yukawa-like interaction between one fermionic and bosonic degree of freedom and consider quantum fluctuations induced by the self interaction the bosons. Within the framework of the functional renormalization group (FRG) method, using the  Wetterich equation, we can compute thermodynamic quantities in the Local Potential Approximation (LPA) with the optimized Litim regulator by~\cite{Litim:2001}. The obtained equation of state is investigated by solving the Tolman\,--\,Oppenheimer\,--\,Volkov (TOV) equations to calculate the mass-radius relation of compact stars at different levels of approximation for the quantum fluctuations. We have also calculated the effect of quantum fluctuations on compressibility and compactness. This latter is an important quantity to predict gravitational waveform parameters originating from e.g. binary neutron star mergers~\cite{Hinderer:2009ca}. Our aim with this demonstrative model is to examine the effect of quantum fluctuations on astrophysical observable quantities and to study how and why these effect emerge. Here, we predict the magnitude of the uncertainty of macroscopical observables caused by microscopical fluctuations. 

\section{Bosonic fluctuations in the interacting Fermi gas model}

To take into account the effect of quantum fluctuations we use the functional renormalization group method, which is a general way to calculate the effective action of a system in quantum field theory. The obtained effective action contains all quantum effects and can be used directly to calculate the thermodynamics of the system because it only contains measurable, infrared safe (IR) quantities. 

The FRG method introduces a regulator term $R_{k}$ in the action, which acts as a mass term, and suppresses modes below the momentum scale $k$. Varying this scale, the quantum fluctuations are taken into account at a certain scale level. In our calculations we use the optimized Litim regulator~\cite{Litim:2001}, which minimizes the regulator-choice dependence of the results. This modification makes the effective action $\Gamma_{k}$ scale dependent, which evolution is described by the Wetterich equation as explained in Refs. by~\cite{Wetterich:1989xg,Gies:2006wv,Wetterich:1992yh},
\begin{equation} 
\partial_{k} \Gamma_{k}=\frac{1}{2} \, \int \dd p^D \,  {\rm STr} \, \left [ \left( \partial_{k}R_{k} \right) \left( \Gamma_{k}^{(2)} + R_{k} \right)^{-1} \right],
\label{wetterich}
\end{equation}
where $\Gamma_{k}^{(2)}$ is the second derivative matrix of the effective action. The term 'STr' denotes the normal \textit{trace} operation but includes a negative sign for fermionic fields and sums over all indices. Integrating the Wetterich equaiton~\eqref{wetterich} at finite chemical potential yields an effective actions which can be used to calculate thermodynamic quantities and the EoS for nuclear matter at zero temperature and high densities. 
Since we are interested in effective theories of nuclear matter, these theories  do not have to be renormalizable. They contain a cut-off which defines the highest energy scale where the theory is still considered as valid. The parameters of the model are choosen in a way that they reproduce the measured properties of the ordinary nuclear matter. The resulting EoS and compact star parameters are the predictions of the model with this initial condition.   

The direct integration of the Wetterich equation~\eqref{wetterich} is not possible in a general way, thus we have to make assumptions for the form of the effective action. To study the effect caused by the quantum fluctuations on the EoS and on macroscopical compact star observables, we use a simple Yukawa-type model with one bosonic and one fermionic degree of freedom described by the bare action:
\begin{equation}
\begin{aligned}
  \Gamma_k[\ph,\psi] &=\\
  = \int & \dd^4x\left[\bar \psi(i\sls \d -g\ph)\psi +
    \frac12 (\d_\mu\ph)^2 - U_k(\ph)\right],
\label{bareaction}
\end{aligned}
\end{equation}
where $\ph$ is the bosonic field of which fluctuations are studied and $\psi$ is the fermionic field corresponding to nucleons in e.g. a more sophisticated model. The effect of bosonic fluctuations is characterized by the scale-dependent effective potential $U_{k}$. We make the \textit{ansatz} that the form of effective action~\eqref{bareaction} does not change during the integration of the Wetterich equation. This is consistent within LPA which is a non-pertrubative \textit{ansatz} and it means that the momentum dependence of the vertices is dropped, but we can extend our calculations to the regions of strong couplings. Following~\cite{Posfay:2015} the Wetterich equation for $U_{k}$ at finite temperature is,
\begin{equation}\label{wetterich2}
\begin{aligned}
 \d_k U_k = {} & \frac12\mathop{\mathrm{STr}} \ln \left[ R_k + \Gamma_k^{(2)} \right] =\\
 = \frac{k^4}{12\pi^2}&\left[ 
    \frac{1+2n_B(\omega_B)}{\omega_B}+ \right.\\
   & + \left. 4\,\frac{-1+n_F(\omega_F-\mu)+n_F(\omega_F+\mu)}{\omega_F}\right] ,
\end{aligned}
\end{equation}
where $n_B$ and $n_F$ are the Bose\,--\,Einstein and the Fermi\,--\,Dirac distributions, respectively. Both are defined by 
\begin{equation}
\begin{aligned}
 & n_{B/F}(\omega)= {} \frac1{e^{\beta\omega}\mp 1} \\ & \omega_B^2 = k^2 +\d_\ph^2 U, \\  & \omega_F^2 = k^2 +g^2\ph^2,
\end{aligned}
\end{equation}
where $g$ is the coupling, and $\beta =1/T$ is the inverse temperature of the system.
Note in this method the thermodynamical parameters enter through the flow equation at finite temperature and chemical potential.

At the cutoff scale we choose the initial conditions for the potential in the form of $\varphi^4$ model: 
\begin{equation}
U_{\Lambda}=U_{k=\Lambda}=\frac{m_{\Lambda}^2}{2} \varphi^2 + \frac{\lambda_{\Lambda}}{24} \varphi^4.
\end{equation} 
The bare parameters $m_{\Lambda}$ and $\lambda_{\Lambda}$ were choosen in a way that our obtained equation of state agrees  with other  phenomenologically more sophisticated EoS at high energies~\cite{sqmproc:2017}.
The specific Wetterich equation~\eqref{wetterich2} for the demonstrative model given by~\eqref{bareaction} is solved by the technique proposed by~\cite{harmonic:2017} at zero temperature and finite chemical potential. The cut-off scale is chosen to be $\Lambda=1.4\,\text{GeV}$ as a typical value applied in a similar model by~\cite{Drews:2014spa}. The obtained EoS were studied in Ref.~\cite{sqmproc:2017} and the corresponding neutron star parameters were compared to other models. In this work we show how the effect of quantum fluctuations manifest themselves in the observable quantities of compact stars when considering different levels of approximations. 

\section{Compact star observables and quantum fluctuations}

In order to constrain the equation of state of the dense nuclear matter the most important measurable properties of the neutron stars are their mass, $M$ and radius, $R$. The mass-radius curve shows the possible neutron star configurations of a given EoS. In the case of our demonstrative model, given by the action~\eqref{bareaction}, we calculated the mass-radius diagram for different levels of approximations in~\cite{sqmproc:2017}.

\subsection{The effect of fluctuations on compactness}

Merging compact stars are also predicted sources of gravitational waves~\cite{Rezzolla:2016nxn}. The form of the emitted gravitational waves strongly depends on the compactness of the neutron stars, which is defined as $C={M}/{R}$~\citep{Hinderer:2009ca}. We calculated here how the compactness is related to the mass and the radius of the neutron star in our toy model to study the effect of quantum fluctuations. We compared our model to some other EoS taken from Ref.~\cite{Ozel:2016}, which are typically used in compact star models. The physical quantities are calculated in different levels of approximation: the mean field level which contains no quantum effects, the one-loop level which is the lowest order approximation for quantum effects, and the high-order FRG LPA approximation, which is the most accurate from these ones. 

\begin{figure}[h]
\begin{center}
\includegraphics[width=\columnwidth]{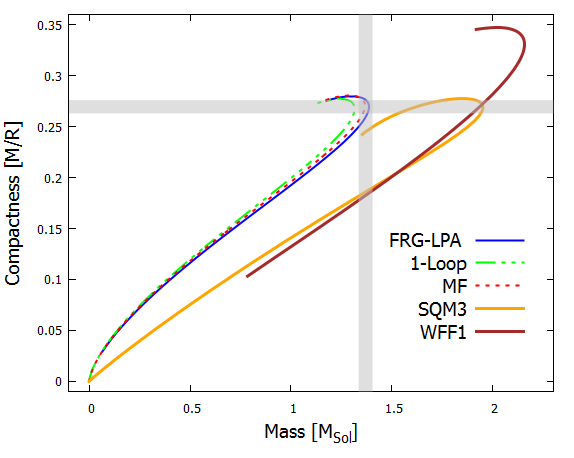}
\caption{The compactness-mass relation of neutron stars calculated within the FRG-framework in different approximations. Comparison to models WFF1 and SQM3 are also presented, including the uncertainty of the observational resolution by shaded bars~\cite{Ozel:2016}.}\label{mc}
\end{center}
\end{figure}

\begin{figure}[h]
\includegraphics[width=\columnwidth]{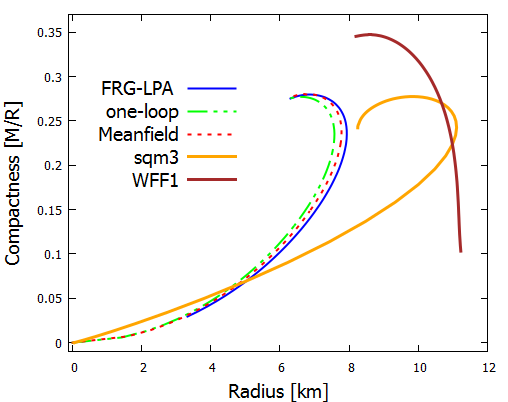}
\caption{The compactness-radius relation of neutron stars calculated within the FRG-framework in different approximations. Comparison to models WFF1 and SQM3 are also presented~\cite{Ozel:2016}.}\label{rc}
\end{figure}

Relations are shown on Figs.~\ref{mc} and~\ref{rc} are important when the neutron star parameters has to be determined from gravity waves. The compactness is related to the waveform but the equation of state is needed for the mass and/or radius estimation. The obtained curves present, that considering quantum fluctuations in different approximations changes the predicted value of mass and radius -- corresponding to a neutron star with a given compactness. The deviation between calculations is the most prominent in the domain of the large-mass and high-density compact stars which are the most relevant cases from both nuclear physics and astrophysics points of view~\cite{Kojo:2017pfw}. 

The maximal deviation between approximations is about $5\, \%$ for mass and radius. The FRG-LPA method always predict larger mass neutrons stars than the other two. The mean field approximation is closer to the high order FRG-LPA approximation, which shows that taking into account high-orders are needed in the FRG-LPA method for modeling the dense matter. 

Based on these calculations one can argue, that using mean field models is enough to study the neutron star data, since it is so close to the high order calculation. However, one has to be careful with such a prediction because as shown by~\cite{harmonic:2017} the phase diagram of the model behaves differently in various approximations. The analyzed phase structure shows about $30\%$ difference between the mean field and high order calculations. This fact becomes important if one wants to use astrophysical data to study the phases of dense nuclear matter, for example in studies of color flavour locking (CFL) or  color superconductor phase -- where the phases are quite close to each other. Moreover these results shows that including quantum fluctuations are important when the considered model has more phases and the effect of quantum fluctuations depend on the interactions in the model and they can change the mass and radius of the resulting neutron star. 

So far there is no known compact star with high-precision measurement including both mass {\sl and} radius, thus one should use data from various sources for uncertainty estimates.
On Figure~\ref{mc}, an uncertainty estimate of the recent observational data was plotted by {\sl shaded bars}. Note, that these uncertainty bars were taken based on~\cite{Ozel:2016}, however we note mass-only measurement alone can be more accurate, due to the Shapiro-delay based precise measurements~\cite{Demorest:2010bx,vanKerkwijk:2010mt}.

Using the demonstrative model described in~\cite{harmonic:2017} we can estimate whether or not the effect of quantum fluctuations on compact star observables can be detected by modern data analysis methods. For the estimation we assume that the relative size of the effect of fluctuations remains the same in more realistic models of neutron stars. We also consider the most optimistic case on Figure~\ref{mc} where the effect of fluctuations is the biggest.Comparing the experimental uncertainty to the effect of fluctuations, one can see that they are in the same order of magnitude, although the accuracy of the measurements is not yet enough to provide constraints based on the effect of fluctuations.

\subsection{Compressibility-sensitivity on fluctuations}

The compressibility of nuclear matter is one of the most sensitive measure of the nuclear matter. It is also strongly influences the mass and radius of the resulting neutron star, and related to compactness too. The compressibility, $\chi$, is defined by 
\begin{equation}
\frac{1}{\chi}=n \, \frac{\partial P}{\partial n},
\label{termcomp}
\end{equation}
where $n$ is the density of matter and $P$ is the pressure. The compressibility given by eq.~\eqref{termcomp} can also be expressed as a function of average energy per particle:
\begin{equation}
\frac{1}{\chi}=2 n^2 \frac{\partial }{\partial n} (E/A)+
n^3 \frac{\partial^2 }{\partial n^2}(E/A),
\label{longcomp}
\end{equation}
where $E/A$ is the energy of one nucleon. If the $E/A$ has a minimum, the model describes a stable nucleus with binding energy, hence the saturation density $n_{0}$ and the compression modulus $K$ is defined at the minimum: 
\begin{equation}
K=k_{F}^2\frac{\partial^2 }{\partial k_{F}^2}(E/A)
\end{equation}
where $k_{F}$ is the Fermi-momentum. In this case the compressibility and the compression modulus are connected:
\begin{equation}
K=\frac{9}{n_{0} \, \chi} \ .
\end{equation}

However, our demonstrative model has no binding energy, but the compressibility can be calculated. In a more sophisticated model there can be only a multiplicative constant difference between the compressibility and the compression modulus. Comparing compressibility as a function of relative density at different levels of approximation implies that the fluctuations has a similar effect on the compression modulus in more sophisticated models.

\begin{figure}[h]
\includegraphics[width=\columnwidth]{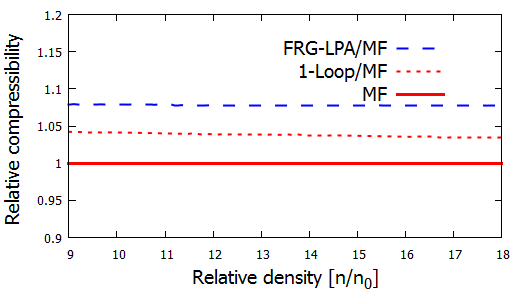}
\caption{Compressibility calculated at different approximation of the EoS, as a function of nuclear density. Everything is normalized to the mean field approximation values.}\label{comp}
\end{figure} 

The results are shown on Fig.~\ref{comp} relative to the mean field calculations as a reference ({\sl solid line}). Since our model is a proof-of-concept demonstrative-model and does not contain a repulsive force except the Fermionic-nature of nucleons, the densities are higher than the nuclear saturation density $n_{0}=0.153$~fm\textsuperscript{-3}. Due to this, the compressibility values are also higher, but here we are interested only in the relative difference of the corresponding approximations. 

For this reason on Fig.~\ref{comp} we show the relative compressibility compared to the mean field calculation, which constraints no quantum effects. The high order FRG calculation ({\sl dashed line}) is the most accurate and it is approximately $8\,\%$ higher then the mean field result. The one-loop calculation, which considers quantum fluctuations at the lowest order ({\sl dotted line}) is about $3-4\, \%$ higher than the reference mean field one. Fig.~\ref{comp} shows that the first order approximation is not satisfying for the purposes of calculating the compressibility. 
 

\section{DISCUSSION}

Taking into account the recent experimental devices, analysis methods, and the expected statistics, we 
compared the obtained uncertainty to the calculated effect of the quantum fluctuations. As Figures~\ref{mc},~\ref{rc}, and~\ref{comp} show, the proposed difference originated by taking into account the quantum fluctuations at different order, appears as $\lesssim 5\%$ effect on the macroscopical observables of compact stars. In comparison to this, the experimental data were found to have about $\lesssim 10\%$ used Refs.~\cite{Demorest:2010bx,Ozel:2016}. 

\begin{figure}[h]
\includegraphics[width=1.1\columnwidth]{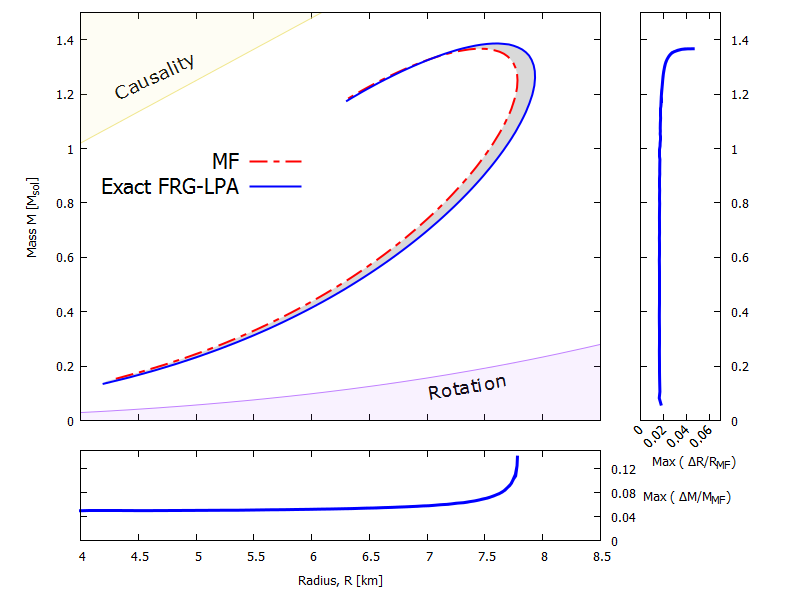}
\caption{The $M(R)$ diagram calculated for the FRG-method (FRG-LPA) and mean field (MF) approximation, including maximum relative deviation caused by the quantum fluctuation effects on mass and radius.}\label{mrdiff}
\end{figure}

Certainly, our comparison presents the best-case scenario, especially, in sense of the theoretical calculations, where the maximal deviance was taken into account. To quantify this, on Fig.~\ref{mrdiff} we presented the mass-radius relation including the difference of the FRG-LPA $M(R)$ result relative to the mean field approximation as {\sl shaded area}. 

Based on our investigation for the maximal relative difference on  mass and radius, $\max(\Delta M/M_{MF}(R))$ and $\max(\Delta R /R_{MF}(M))$ respectively, which are plotted on the {\sl bottom} and {\sl side} graphs of the Fig.~\ref{mrdiff}, one can see easily, that the relative deviation is the largest at the maximal mass and radius close to the stable solution. Here $\max(\Delta M /M_{MF}(R_{max})) \ge 0.1$ and  $\max(\Delta R /R_{MF}(M_{max}) \ge 0.05$. The difference between the mean field and FRG-LPA calculations is approximately constant for other sections of the M-R diagram and approximately  $4\%$ for the mass and $2 \%$ for the radius. 

It is important to emphasize that our model is demonstrative, and the deviation on the M-R diagram may vary according to the interplay of interaction terms in other models, however the deviations proposed to have the same order of magnitude as in our model.   

\section{CONCLUSION}

Based on Refs.~\cite{Posfay:2015,harmonic:2017,sqmproc:2017} using the FRG method, we calculated the effect of bosonic quantum fluctuations on neutron star observables considering the simplest interacting Fermi-gas model and compared our results to some other similar model EoS as well. We concluded that high-order calculations are needed for the consistency between the phase diagram and the observable quantities such as the mass and radius of compact stars, compactness, and thermodynamical quantities like compressibility. 

We presented that the difference between compact star observables corresponding to the various approximations of the EoS is in the order of $\sim 5\%$. The effect is most prominent on the M-R diagram near the maximum mass and radius. We have also concluded that although modern measurements are not accurate enough to detect these effects, they are very close and it is probable that high-statistics and high accuracy NICER data analyzed by novel techniques will resolve the deviations caused by quantum fluctuations.

\begin{acknowledgements}
This work was supported by Hungarian National Funds (OTKA) grants
K120660 and K123815. NewCompStar MP1304, PHAROS CA16214, and THOR
CA15213 COST actions and NKM-81/2016 MTA-UA bilateral mobility
program.
\end{acknowledgements}

\bibliographystyle{pasa-mnras}
\bibliography{myreferences2}

\end{document}